\documentclass[12pt]{article}

\usepackage{amssymb, amsmath}
\usepackage{graphicx}
\usepackage{cite}

\def\be{\begin{equation}}
\def\ee{\end{equation}}
\def\bea{\begin{eqnarray}}
\def\eea{\end{eqnarray}}

\newcommand{\td}{\text{d}}
\usepackage{hyperref}
\usepackage[left=2cm,right=2cm,top=2cm,bottom=2cm]{geometry}

\title{ \bf{On the nonexistence of extreme anti-de Sitter black rings }}

\author{James Lucietti\footnote{j.lucietti@ed.ac.uk } 
\\ \\ \small \sl School of Mathematics and Maxwell Institute for Mathematical Sciences, \\ \small \sl    University of Edinburgh, King's Buildings, Edinburgh, EH9 3FD, UK }

\date{}

\begin{document}

\maketitle

\begin{picture}(0,0)(0,0)
\put(350, 240){}
\put(350, 225){}
\end{picture}

\begin{abstract}  
We prove that five-dimensional extreme anti-de Sitter black ring solutions to the vacuum Einstein equations that admit biaxial symmetry do not exist.  
This is established by demonstrating the nonexistence of five-dimensional, biaxisymmetric, vacuum near-horizon geometries of extreme horizons with a nonzero cosmological constant and ring topology. 
\end{abstract}

\vspace{1cm}

A striking result in higher dimensional General Relativity is the failure of the black hole uniqueness theorem~\cite{Emparan:2008eg}. This was revealed by the existence a black ring solution to the five-dimensional vacuum Einstein equations, i.e. an asymptotically flat black hole solution with a horizon of $S^1\times S^2$ topology~\cite{Emparan:2001wn}.  A basic question is whether such black holes exist in the presence of a cosmological constant. Approximate solutions have been constructed corresponding to  thin black rings in anti-de Sitter and de Sitter spacetime, for which the radius of the $S^2$ is much smaller than the cosmological scale~\cite{Caldarelli:2008pz}.  Furthermore, anti-de Sitter black rings have been constructed numerically in regimes not accessible to such approximations~\cite{Figueras:2014dta}. Curiously, these works indicate the nonexistence of thin anti-de Sitter black rings whose $S^2$ is much larger than the cosmological scale, in agreement with expectations from the AdS/CFT correspondence (i.e. there are no corresponding fluid configurations in the hydrodynamic regime of the CFT~\cite{Bhattacharyya:2007vs}). 

There are also a number of nonexistence theorems. It has been shown that supersymmetric anti-de Sitter black ring solutions to minimal gauged supergravity do not exist (Einstein-Maxwell-$\Lambda$ theory coupled to a Chern-Simons term)~\cite{Kunduri:2006uh, Grover:2013hja}.  Even more surprising, it has recently been proven that extreme de Sitter black rings, possibly coupled to matter obeying the dominant energy condition, do not exist~\cite{Khuri:2017zqg, Khuri:2018rzo}. These nonexistence results rely on the fact that for any extreme black hole solution the near-horizon geometry also satisfies the Einstein equation and so horizon topologies and geometries can be classified (and ruled out) independently of the black hole classification problem, see~\cite{Kunduri:2013ana} for a review.

In this note we will consider five-dimensional spacetimes that obey the vacuum Einstein equations with a cosmological constant $\Lambda$ 
 and which contain an extreme (degenerate) Killing horizon. As is well known, the Einstein equation for the near-horizon geometry then reduces to the following  geometric equation for the Riemannian metric $g_{ab}$ induced on a three-dimensional cross-section $S$ of the horizon 
\be
R_{ab} = \tfrac{1}{2} h_a h_b - D_{(a} h_{b)} + \Lambda g_{ab} \label{nheq}
\ee
where $D_a$ is the metric connection of $g_{ab}$  and $h_a$ is a 1-form on $S$. For applications to the study of black holes $S$ is assumed to be a compact manifold (no boundary). The classification of solutions to this horizon equation has been extensively studied~\cite{Kunduri:2013ana}. For $\Lambda=0$ a complete classification was derived for solutions that admit $U(1)^2$-symmetry~\cite{Kunduri:2008rs}. The assumption of such biaxial symmetry is compatible with both asymptotically flat and Kaluza-Klein (KK) spacetimes, so this classification captures all the known extreme black hole solutions  (Myers-Perry black holes, black rings, KK black holes, black strings) and also revealed there are no other extreme horizons in this class. 

The $\Lambda \neq 0$ case was also considered and the classification reduced to a 6th order nonlinear ODE of a single function~\cite{Kunduri:2008rs}. Other than the known rotating black hole~\cite{Hawking:1998kw}, no other solutions to this ODE were found, so the results in this case were inconclusive.  Of course, with hindsight,  we now know that for $\Lambda>0$ there can be no black ring solutions from the recent topology theorems~\cite{Khuri:2018rzo, Khuri:2017zqg}. The purpose of this work is to revisit the $\Lambda \neq 0$ case and point out that there is in fact an elementary proof of the nonexistence of black rings that admit $U(1)^2$-symmetry valid for both $\Lambda >0$ (de Sitter) and $\Lambda <0$ (anti-de Sitter). In particular, our main result is that five-dimensional extreme vacuum anti-de Sitter black rings that admit biaxial symmetry do not exist.\footnote{It is worth emphasising that our result does not invoke any global assumptions on the spacetime and hence is valid for both asymptotically AdS and locally AdS spacetimes.} 

We first need to recall the relevant results of~\cite{Kunduri:2008rs}. Suppose $S$ is a three-dimensional compact (nontoroidal) manifold with $U(1)^2$-symmetry and let $\eta_i, i=1,2,$ denote the commuting  Killing fields which generate the biaxial symmetry.  Then, the orbit space $S/U(1)^2$ is a closed interval and the matrix $\gamma_{ij} \equiv g(\eta_i , \eta_j)$ is rank-1 at the endpoints and rank-2 in the interior (see e.g.~\cite{Hollands:2007aj, Hollands:2009ng}). The topology of $S$ is then determined by the null space of $\gamma_{ij}$ at the two endpoints:  $S^1 \times S^2$ (ring) if the null spaces at the endpoints are the same;  or  locally $S^3$ otherwise (the latter includes the lens spaces). Now, there exists a globally defined $U(1)^2$-invariant function $x$ such that $\td x = - i_{\eta_1} i_{\eta_2} \epsilon_g$ where $\epsilon_g$ is the volume form of $g_{ab}$. Compactness of $S$, together with the relation $| \td x |^2 = \det \gamma_{ij}$,  implies that $x$ attains precisely one minimum and one maximum, say $x_0$ and $x_1$ respectively. Hence,  we may use $x$ as a coordinate on the interior of the orbit space and in particular identify the orbit space $S/U(1)^2 \cong [ x_0, x_1]$. 

Next, one can decompose the 1-form $h = \beta+\td \lambda$ globally on $S$, where $\beta$ is a co-closed 1-form and $\lambda$ is a function which are $U(1)^2$-invariant.  One then may use coordinates $(x, \phi^i)$ on $S$ adapted to the symmetry so $\eta_i = \partial_{\phi_i}$ and show that $\beta = \beta_i (x) \td\phi^i$ and
\be
g = \frac{\td x^2}{\det\gamma_{ij}} + \gamma_{ij}(x) \td\phi^i \td\phi^j  \; .
\ee
These coordinates are valid for $x_0<x<x_1$ and break down at the endpoints where $\det \gamma_{ij} = 0$. In order that the metric extends smoothly onto a compact manifold $S$ we must require $(\td |v_I|)^2  \to 1$ at each endpoint $x\to x_I$, $I=0,1$, where $v_I= v_I^i\eta_i$ is the vector which vanishes at $x=x_I$ normalised to be $2\pi$-periodic (i.e. $v_I^i$ is in the null space of $\gamma_{ij}(x_I)$). By an $SL(2, \mathbb{Z})$ transformation we may always set $v_I= \eta_1$ for one $I$, so $\gamma_{i 1}(x_I)=0$ and $\gamma_{22}(x_I)>0$. Then, in terms of the the proper distance $s$ from $x_I$, $s=\int_{x_I}^x \sqrt{g_{xx}} \td x$ (which is a global coordinate on $[x_0, x_1]$), smoothness requires $\gamma_{11}= s^2+O(s^4)$,  $\gamma_{12}= O(s^2)$ and $\gamma_{22}=\gamma_{22}(x_I)+O(s^2)$ as $s \to 0$. It follows that $\sqrt{\det \gamma_{ij}}$  as a function of $s$ has simple zeros at the endpoints.

It proves convenient to introduce the globally defined $U(1)^2$-invariant functions $\Gamma \equiv e^{-\lambda}$ and $Q \equiv \Gamma \det \gamma_{ij}$. Note that $\Gamma>0$ everywhere and $Q(x) >0$ for $x_0<x<x_1$ and vanishes at the endpoints $Q(x_0)=Q(x_1)=0$.  
Smoothness then implies that $Q(x)$ as a function of $x$ has simple zeros and thus, since $Q(x)>0$ in the interior, in particular
\bea 
 Q'(x_0)>0, \quad Q'(x_1)<0  \; .  \label{BC}
\eea
This is not obvious since $x$ is not a well defined coordinate at the endpoints. 
 To see it, note that in terms of the proper distance $s$ we have $\td x/\td s = \sqrt{\det \gamma_{ij}}$ which, as noted above, has simple zeros at the endpoints. It follows that $x-x_I$  has a double zero at the endpoint $x_I$ and also that $Q$ has double zeros at the endpoints. Therefore, $Q(x)$ as a function of $x$  has simple zeros, as claimed.

In~\cite{Kunduri:2008rs} it was shown that, for any solution to (\ref{nheq}) of the above form, the pair of functions $(\Gamma(x), Q(x))$  obey the ODE system
\bea
&&\frac{\td}{\td x} \left( \frac{Q^3}{\Gamma} \frac{\td^3 \Gamma}{\td x^3}\right) = 10 \Lambda Q^2 \frac{\td^2\Gamma}{\td x^2}  \label{Gameq} \\
&&\frac{\td^2 Q}{\td x^2} + 2 C^2 + 6 \Lambda \Gamma=0 \label{Qeq}
\eea
where $C>0$ is a constant (this constant can be set to any value by certain scalings).  Furthermore, the full solution may be reconstructed from this data:
\be
g = \frac{\Gamma}{Q} \td x^2 + P (\td\phi^1+ \omega \td\phi^2)^2 + \frac{Q}{\Gamma P} (\td\phi^2)^2  \; ,
\ee
where the other metric components are determined by 
\bea
P &=& \Gamma \frac{\td}{\td x} \left( \frac{Q \Gamma'}{\Gamma} \right) + 2 \Gamma (C^2+ \Lambda \Gamma)  \label{P}\\
P^2 \omega'^2 &=& \frac{1}{\Gamma} \frac{\td}{\td x} \left( \frac{Q P'}{P} \right)+ 2\Lambda + \frac{P}{\Gamma^2} \label{omega1}
\eea
and $f' \equiv \td f/\td x$.
It was also shown that 
\be
\omega' = \frac{k}{P^2 \Gamma}  \label{omega3}
\ee
where $k$ is a constant.  Note that to derive the above a $GL(2, \mathbb{R})$ transformation  $\eta_i \to A_{ij}\eta_j$ was performed to align $k^i \equiv \Gamma \gamma^{ij} \beta_j= \delta^i_1$ (constancy of $k^i$ follows from the $(xi)$ component of (\ref{nheq})). Thus, in these coordinates the $\eta_i$ may not have closed orbits.  The boundary conditions for the above ODE system at the endpoints $x=x_I$ are fixed by the requirement that the metric extends to a smooth metric on a compact $S$, as discussed above.  A key consequence of this, which we will use below, is that for any smooth $U(1)^2$-invariant function $f$ on $S$ the function $Qf' =\Gamma \sqrt{\det\gamma_{ij}}\frac{\td f}{\td s}$ is smooth and vanishes at the endpoints (recall $s$ is the proper distance as above).

For $\Lambda=0$, the system is easy to solve~\cite{Kunduri:2008rs}. Then, $Q^3 \Gamma'''/\Gamma$ is a constant and evaluating this at the endpoints implies this must vanish everywhere.\footnote{Again, since $x$ is not a valid coordinate at the endpoints, one must take care here. One can see that $Q^3 \Gamma'''$ vanishes at the endpoints by writing it in terms of the proper distance $s$.} Thus $\Gamma'''=0$, so $\Gamma$ and $Q$ are both quadratic functions of $x$ and the horizon metric is fully determined, leading to a complete classification. 
We wish to solve the $\Lambda \neq 0$ system.  Clearly $\Gamma(x)$ linear solves the system; this gives the known rotating black hole with a cosmological constant~\cite{Kunduri:2008rs}.  But are there any other solutions?  As mentioned above, for $\Lambda>0$ it has been argued on general grounds that there can't be any ring topology horizons~\cite{Khuri:2017zqg}. However, the situation for $\Lambda<0$ has remained unclear.

We now turn to the $\Lambda\neq 0$ case.
Observe that $P=| \eta_1 |^2$ and hence $P \geq 0$ and may vanish only at the endpoints. 
There are a number of cases depending on which Killing field vanishes at the endpoints.  First suppose  $P>0$ everywhere on $S$. Then the Killing field
\be
v_I = \eta_2 - \omega(x_I) \eta_1
\ee
vanishes at $x=x_I$ for $I=0,1$, respectively. If $k\neq 0$ equation (\ref{omega3}) implies $\omega$ is a monotonic function so that $\omega(x_0)\neq \omega(x_1)$. Thus $v_0 \neq v_1$ and hence the topology in this case must be locally $S^3$. Therefore, for ring topology we must have $k=0$ and hence $\omega'=0$.  Then (\ref{omega1}) gives
\be
\frac{\td}{\td x} \left( \frac{Q P'}{P} \right)+ 2\Lambda \Gamma + \frac{P}{\Gamma}=0  \
\ee
and integrating this over the interval we find the boundary term vanishes giving
\be
\int_{x_0}^{x_1} \frac{P}{\Gamma} \td x = - 2\Lambda \int_{x_0}^{x_1}  \Gamma  \td x  \; .  \label{int1}
\ee
Notice for $\Lambda\geq 0$ we immediately get a contradiction. For $\Lambda<0$ we need to work a little harder. Integrating (\ref{P}) divided by $\Gamma$ gives
\be
\int_{x_0}^{x_1} \frac{P}{\Gamma} \td x = \int_{x_0}^{x_1} ( 2C^2 +2 \Lambda \Gamma) \td x
\ee
and equating this with (\ref{int1}) gives
\be
 \int_{x_0}^{x_1} ( C^2 +2 \Lambda \Gamma) \td x=0   \; .\label{Pposcase}
\ee
On the other hand, integrating (\ref{Qeq}) gives
\be
Q'(x_1)- Q'(x_0) =- 2 \int_{x_0}^{x_1} (C^2 + 3 \Lambda \Gamma) \td x  =- 2\Lambda \int_{x_0}^{x_1}  \Gamma  \td x
\ee
where in the second equality we used (\ref{Pposcase}).  From the boundary conditions (\ref{BC}) the LHS is negative. However, for $\Lambda \leq 0$ the RHS is nonnegative, giving a contradiction. We conclude that there are no solutions for {\it any} $\Lambda$ with ring topology in the case $P$ is strictly positive (this is consistent with the $\Lambda=0$ results of~\cite{Kunduri:2008rs}).

On the other hand, now suppose $P$ vanishes at one endpoint. Then, ring topology requires it vanishes at both endpoints (it must be the same Killing field vanishing) so $P(x_0)=P(x_1)=0$, i.e. $\eta_1$ vanishes at the endpoints.  As discussed above, smoothness then requires that $(\td | \eta_1 | )^2  \to c^2$  as $x \to x_I$ for $I=0,1$, where $c$ is a constant related to the periodicity of $\phi_1$. Explicitly, in terms of the proper distance $s$, this smoothness condition is  $\partial_s \sqrt{P} \to \pm c$ which implies $\sqrt{P}$ has simple zeros at the endpoints (just like $\sqrt{Q}$). 
It follows that $P/Q$ is a smooth positive function on $S$.
To treat this case it is thus convenient to define a smooth function $F$ which is strictly positive on $S$ by
\be
P = \frac{Q}{\Gamma F}  \label{F}
\ee
in terms of which the horizon geometry is
\be
g= \frac{\Gamma}{Q} \td x^2 + \frac{Q}{\Gamma F} (\td\phi^1+ \omega \td\phi^2)^2 +F (\td\phi^2)^2
\ee
and equation (\ref{omega1}) becomes
\bea
\Gamma P^2 \omega'^2 &=& Q''- \frac{\td}{\td x} \left( \frac{Q \Gamma'}{\Gamma}+ \frac{Q F'}{F} \right) + 2\Lambda \Gamma+ \frac{P}{\Gamma} \\ &=&  - \frac{\td}{\td x}\left( \frac{Q F'}{F} \right) - 2\Lambda \Gamma  \label{omega2}
\eea
where in the second equality we used (\ref{Qeq}) and (\ref{P}). 
 Now, noting that $\omega= \eta_1 \cdot \eta_2 / | \eta_1 |^2$, we deduce that $\omega$ is a smooth function on $S$ (as shown above, in terms of the proper distance $s$, the denominator has a double zero at the endpoints, whereas the numerator has at least a double zero).   But (\ref{omega3}) can be written as $k= \Gamma P^2 \omega' = P F^{-1} Q \omega'$ and evaluating this at the endpoints thus implies $k=0$. Hence  $\omega'=0$ and so (\ref{omega2}) becomes 
\be
\frac{\td}{\td x}\left( \frac{Q F'}{F} \right) + 2\Lambda \Gamma=0 \; . \label{Feq}
\ee
Integrating this over the interval we find the boundary term vanishes leaving
\be
2\Lambda \int_{x_0}^{x_1} \Gamma  \td x =0  \; .
\ee
But by definition $\Gamma$ is strictly positive, so we have found a contradiction unless $\Lambda=0$. Thus there are no $\Lambda \neq 0$ solutions with ring topology in this case either.  For $\Lambda=0$ equation (\ref{Feq}) implies $QF'/F$ is a constant and evaluating this at the endpoints shows that $F$ is a constant; upon removing the conical singularities at the endpoints the resulting regular solution is isometric to a boosted extreme Kerr string horizon (which includes the extreme black ring horizon)~\cite{Kunduri:2008rs}.

Therefore, we have proved that there are no extreme horizons with $S^1\times S^2$ topology and $\Lambda \neq 0$, in particular excluding the anti-de Sitter case $\Lambda<0$.  It is worth noting that our proof also rules out the possibility of an extreme AdS/dS black ring held in equilibrium by a conical singularity, i.e., the above proof remains valid even if one can not simultaneously remove the conical singularities at the endpoints $x=x_I$.\footnote{More precisely, if $v$ is the vector which vanishes at the endpoints, then $(\td |v| )^2 \to (c_I)^2$ as $x\to x_I$ where $c_0- c_1 \neq 0$ is proportional to the angular deficit.} In contrast, for supersymmetric AdS solutions there is a ring horizon with a conical singularity~\cite{Kunduri:2006uh}.

It would be interesting to complete the classification of locally $S^3$ extreme horizons in the $\Lambda \neq 0$ case; a natural conjecture is that the known solution (corresponding to linear $\Gamma(x)$) is the most general.  Furthermore, given the absence of both vacuum and supersymmetric anti-de Sitter black rings, it is now tempting to conjecture the absence of generic extreme charged anti-de Sitter black rings in Einstein-Maxwell type theories (in particular including minimal gauged supergravity), at least with biaxial symmetry.
\\

\noindent {\bf Acknowledgments.} The author is supported by STFC [ST/L000458/1].

\end{document}